\documentclass[preprint,superscriptaddress,amsmath,amssymb,11pt,aps,showpacs,showkeys,nofootinbib]{revtex4}
\usepackage{graphicx}
\usepackage{graphics}
\usepackage{amsmath}
\usepackage{dcolumn}
\usepackage{amssymb}
\usepackage{bm}
\usepackage[latin1]{inputenc} 
\usepackage{epsfig}
\usepackage{hyperref}
\usepackage{color}
 \usepackage{array}

\def\al{\alpha} 
\def\be{\beta} 
\def\ga{\gamma}

\def\ep{\epsilon}

\def\th{\theta}

\def\la{\lambda}

\def\si{\sigma}

\def\De{\Delta}

\def\pa{\partial}
\def\half{\frac{1}{2}}

\newcommand{\ben}{\begin{equation}}
\newcommand{\een}{\end{equation}}
\newcommand{\bea}{\begin{eqnarray}}
\newcommand{\eea}{\end{eqnarray}}
\newcommand{\ba}{\begin{array}}
\newcommand{\ea}{\end{array}}
\newcommand{\bit}{\begin{itemize}}
\newcommand{\eit}{\end{itemize}}

\newcommand{\mH}{m_h}
\newcommand{\vEW}{v_\text{ew}}
\newcommand{\bFT}{\beta_\text{ft}}

\newcommand{\rhoInf}{\rho_\infty}
\newcommand{\thW}{\th_\text{w}}
\newcommand{\tilV}{\tilde{V}}
\newcommand{\tilla}{\tilde{\la}}
\newcommand{\fNO}{\bar{f}}
\newcommand{\aNO}{\bar{a}}

\begin{document}
\newcommand{\Sussex}{\affiliation{
Department of Physics and Astronomy,
University of Sussex, Falmer, Brighton BN1 9QH,
U.K.}}

\newcommand{\HIPetc}{\affiliation{
Department of Physics and Helsinki Institute of Physics,
PL 64 (Gustaf H\"{a}llstr\"{o}min katu 2),
FI-00014 University of Helsinki,
Finland
}}

\newcommand{\JPess}{
\affiliation{Departamento de F\'isica, Universidade Federal da Para\'iba, Caixa Postal 5008, 58051-970, Jo\~ao Pessoa, PB, Brazil}
}

\title{Big-Bang Nucleosynthesis and Gamma-Ray Constraints on Cosmic Strings with a large Higgs condensate}

\author{H. F. Santana Mota}
\email{hm288@sussex.ac.uk}
\Sussex
\author{Mark Hindmarsh}
\email{m.b.hindmarsh@sussex.ac.uk}
\Sussex
\HIPetc

\begin{abstract}
We consider constraints on cosmic strings from their emission of Higgs particles, 
in the case that the strings have a Higgs condensate with amplitude of order the string mass scale, assuming  
that a fraction of the energy of condensate can be turned into radiation near cusps.
The injection of energy by the decaying Higgs particles affects the light element abundances predicted by standard 
Big-Bang Nucleosynthesis (BBN), and also contributes to the Diffuse Gamma-Ray Background (DGRB) in the universe today. 
We examine the two main string scenarios (Nambu-Goto and field theory), and 
find that the primordial Helium and Deuterium abundances strongly constrain the string tension and the efficiency of the emission process in the NG scenario, while the strongest BBN constraint in the FT scenario comes from the Deuterium abundance. 
The Fermi-LAT measurement of the DGRB constrains the field theory scenario even more strongly than previously estimated from EGRET data, 
requiring that the product of the string tension $\mu$ and Newton's constant $G$ is bounded by $G\mu \lesssim 2.7 \times 10^{-11}\be_\text{ft}^{-2}$, where $\be_\text{ft}^{2}$ is the fraction of the strings' energy going into Higgs particles. 
\end{abstract}


\maketitle

\section{Introduction}
%

Higgs particles can be produced in the early universe by linear topological defects predicted in some gauge theories of elementary particle physics with extra symmetries beyond those of the Standard Model.  These cosmic strings are formed as a result of a spontaneous symmetry breaking at phase transitions in the early universe \cite{VS,hindmarsh}.  In its simplest form, a cosmic string is characterized by its tension 
$\mu$ which is of order $\mu\sim M^2$, where $M$ is the energy scale of the symmetry breaking. Once formed, strings evolve under the own tension, and can intersect and self-intersect, and after reconnection, create loops.  The loops oscillate and decay, either into massive radiation of the fields from which the string is made, or into gravitational radiation.  The relative proportion is highly uncertain, for reasons explained in Ref.~\cite{Hindmarsh:2011qj}.  

These two decay channels motivate two scenarios for string evolution: the field theory (FT) scenario based on direct numerical simulations of strings in the Abelian Higgs model \cite{Bevis:2006mj, Hindmarsh:2008dw}, and the Nambu-Goto (NG) scenario 
(see e.g. \cite{VS,hindmarsh,Copeland:2011dx} for reviews), which 
assumes that strings can be treated as infinitely thin, with tension equal to mass per unit length.
Numerical simulations of Nambu-Goto strings in an expanding universe have also been carried out 
\cite{Albrecht:1989mk,Bennett:1989yp,Allen:1990tv}, and there has been significant progress in the understanding of the loop size distribution \cite{Ringeval:2005kr,Olum:2006ix,BlancoPillado:2011dq,Blanco-Pillado:2013qja}.
In the NG scenario, massive radiation is neglected, except at cusps - points where the string doubles back on itself - and kinks - discontinuous points in the tangent vector along the string. Particles can be emitted from cusps either through annihilation of oppositely oriented string segments (cusp annihilation) or by linear classical radiation.

In \cite{Vachaspati:2009kq} it was pointed out that the string fields are generically coupled to the SM Higgs (via the so-called Higgs portal), which leads to the string developing a Higgs condensate in its core, extending a distance $\mH^{-1}$.  Hence one can generically expect strings to decay into Higgs particles: either as part of non-perturbative massive radiation process visible in field theory simulations, or by cusp emission in the NG scenario.
Recent calculations of the power from kink emission differ from each other by orders of magnitude \cite{Lunardini:2012ct,Long:2014mxa}, and we neglect it here pending the resolution of the issue. 

In \cite{Vachaspati:2009kq} it was assumed that the expectation value of the Higgs in the core of the string was set by the string scale $M$. However, a more recent numerical investigation found that the expectation value of the Higgs in the core is of the same order as the electroweak scale $\vEW$ \cite{Hyde:2013fia}. We note that whatever the value of the condensate, the string is not superconducting \cite{Witten:1984eb,VS,hindmarsh}: the electromagnetic $U(1)$ symmetry remains unbroken at the core of the string.

In this paper we extend arguments in \cite{Haws:1988ax} to show that the Higgs condensate can indeed be ``large'' (i.e. $M$ rather than $\vEW$) 
in the region of parameter space where the Higgs portal coupling is larger than the self-coupling of the symmetry-breaking scalar field. 
The parameter space explored in \cite{Hyde:2013fia} did not include this region.

In \cite{Vachaspati:2009kq} it was also assumed that there was a linear coupling between the string and the Higgs field, and that this coupling was of order $M$, the expectation value of the Higgs in the core of the string.   In \cite{Hyde:2013fia} it was argued that this linear coupling had to be of order of the Higgs large-distance expectation value, and hence orders of magnitude smaller. 
It was also briefly pointed out that there could be 
a non-perturbative mechanism operating at cusps, where the condensates overlap and interact,  or ``condensate annihilation''.
We consider the implications of this mechanism, and parametrise it according to the fraction of the available energy per cusp lost as Higgs radiation.

Using this model of non-perturbative cusp emission, we derive constraints from the requirement that the decay products of the emitted Higgs do not spoil the predictions of Big-Bang Nucleosynthesis (BBN) for light element abundances \cite{Kawasaki:2004qu,Kawasaki:2004yh,Jedamzik:2006xz,Ellis:2005ii}, or produce a $\ga$-ray flux inconsistent with that measured by Fermi-LAT \cite{Abdo:2010nz,Berezinsky:2011cp}. 
The BBN constraint is relevant for energy injected at cosmic times $10^{-1}\, \mathrm{s}\lesssim t\lesssim 10^{12}\, \mathrm{s} $, while the DGRB constraint applies for times $t\gtrsim 10^{15} \,\mathrm{s}$. We obtain constraints in both the NG and FT scenarios, presenting them in Tables \ref{table2} and  \ref{table3}, and in Figs. 2 and 3.

Particle production by loops of cosmic string in the NG scenario has been considered in many other contexts. The emission of moduli particles was considered in Refs. \cite{Berezinsky:2011cp, Lunardini:2012ct, Damour:1996pv, Sabancilar:2009sq, Babichev:2005qd, Peloso:2002rx} and the emission of Kaluza-Klein particles by cosmic superstrings was analysed in Refs. \cite{Dufaux:2012np,Dufaux:2011da}. 
Emission of particles by loops of ``thick'' string (whose width is TeV scale rather than GUT scale) was studied in \cite{Kawasaki:2011dp}.

A recent paper \cite{Long:2014mxa} studies carefully the perturbative emission of particles from strings with electroweak-scale Higgs condensates, including scalars, vectors and fermions, both one-particle and two-particle emission, and emission from cusps, kinks and kink-kink collisions. 
The total emission rate in the class of model considered in this paper, where the Higgs condensate is of order the high scale $M$, 
and the non-perturbative cusp emission mechanism operates, 
can be significantly higher. 

There are other cosmological and astrophysical constraints on strings \cite{VS,hindmarsh,Hindmarsh:2011qj}. According to recent observation of Cosmic Microwave Background (CMB) by the Planck satellite \cite{Ade:2013xla}, the cosmic string tension is constrained to be $G\mu<1.5\times 10^{-7}$ in the Nambu-Goto string model (NG) and $G\mu<3.2\times 10^{-7}$ in the Abelian-Higgs field theory model (FT) (with 95\% confidence level), where $G$ is the Newton's gravitational constant. 
Adding BICEP2 data reduces the Abelian Higgs 95\% confidence upper limit to $G\mu<2.7\times 10^{-7}$ \cite{Lizarraga:2014xza}.
Previous analysis of CMB based on WMAP and ACT data provides, respectively, $G\mu<4.2\times 10^{-7}$ (FT scenario) \cite{Urrestilla:2011gr} and $G\mu<1.6\times 10^{-7}$ (NG scenario) \cite{Dunkley:2010ge}, also both at $95\%$ confidence level. These constraints are translated into a bound on the energy scale of $M\lesssim 10^{15}$ GeV. 
In the NG model, cosmic strings can also be investigated through the emission of gravitational radiation in a wide range of frequencies \cite{PhysRevD.31.3052,Damour:2000wa, Damour:2001bk, Damour:2004kw}.  For instance, the most recent bounds $G\mu<5.3\times 10^{-7}$ and $G\mu\leq 2.8\times 10^{-9}$ are due to pulsar timing arrays and can be found in Refs. \cite{Sanidas:2012ee} and \cite{Blanco-Pillado:2013qja}, respectively. The differences in the bounds reflect different assumptions about the size distribution of string loops.
Finally, there is a strong bound on the FT scenario from the DGRB. Prior to this work, an old analysis of EGRET data \cite{Bhattacharjee:1998qc} gave an estimate $G\mu \lesssim 10^{-10} f^{-1}$ \cite{Hindmarsh:2011qj}, where $f$ is the fraction of the strings' energy going into $\gamma$-rays.

The paper is organized as follows. In Sec. II we present the rate at which Higgs particles are emitted through non-linear interaction in the NG scenario as well as the rate per unit volume of a network of strings in the FT scenario. In Sec.\ \ref{s:EvoStrLooDis} we describe the loop distribution of strings formed in the radiation era and check that the friction-dominated epoch is long over by the time the constraints are applied. Cosmological constraints on $G\mu$ are obtained in Sec. IV. Finally, in Sec. V our conclusions are presented. We also have dedicated an Appendix for details of the calculation of the size of the Higgs condensate in the string core.

In this paper we use natural units $c=\hbar=1$ so that the Newton's gravitational constant can be expressed as $G=m_P^{-2}=t_P^2$, where $m_P\simeq 1.2\times 10^{19}\,\mathrm{GeV}$ and $t_P\simeq 5.4\times 10^{-44}\, \mathrm{s}$ are the Planck mass and time respectively. 
We also take the present time and the time of equal matter and radiation densities to be $t_0\simeq 4.4\times 10^{17}$\,s and $t_{\mathrm{eq}}\simeq 2.4\times 10^{12}$\,s.
%
\section{HIGGS RADIATION FROM STRINGS}
\label{II}
%
\subsection{The Higgs condensate}
%
If the Higgs field has a suitable interaction with the scalar field which makes the string, it can condense (i.e.\ acquire a vacuum expectation value) in the core of the string  \cite{Witten:1984eb,Vachaspati:2009kq}. Such an interaction is provided by the Higgs portal. If the string is made by a complex field $\phi$, then the most general renormalisable potential including the Standard Model Higgs doublet $\Phi$ can be written 
\ben
V(\phi,\Phi) = \half \la_1 (|\phi|^2  - M^2)^2 + \half \la_2 (|\Phi|^2 - \eta^2)^2 +  \la_3(|\phi|^2 - M^2)(|\Phi|^2 - \eta^2),
\label{Po}
\een
We suppose that $M \gg \eta$, and that $\la_1\la_2 > \la_3^2$, in which case the ground state is $|\phi| = M$ and $|\Phi| = \eta$  \cite{Haws:1988ax}. We recognise $\vEW=\sqrt{2}\eta$ as the electroweak scale 246 GeV.  
In \cite{Haws:1988ax}, models where the Higgs does not interact with the gauge field in the string were studied, and it was shown that 
the Higgs takes an expectation value $M$ in the core of the string, providing $\la_3 \gtrsim \la_1$. Outside the string core, the Higgs field decreases as $1/r$.
A detailed analysis of the slightly more complicated model relevant here, where the Higgs and the string gauge field are coupled, 
is given in the Appendix. It is shown that $|\Phi| \sim M\sqrt{{\la_3}/{\la_2}}$ at the core of string and decreases as $1/\sqrt{\la_2}r$ outside it, eventually decreasing exponentially for distances greater than $\mH^{-1}$.

Recent numerical investigations in \cite{Hyde:2013fia} of the solutions with the above potential did not explore the region of parameter space with \( M \gg \eta \) and $ \la_3 > \la_1$ at the same time. It was found that, outside this region, the Higgs expectation value in the string core was of order $\eta$. 

In \cite{Vachaspati:2009kq} it was argued that there was a linear coupling between the string and the Higgs field, which leads to the formation of a Higgs condensate in the core of the string  with amplitude $M$. Subsequently, an argument was given in \cite{Hyde:2013fia} that the coupling must be proportional to $\eta$ as the string can couple to the Higgs only after electroweak symmetry breaking. However, as is noted by the same authors, the electroweak symmetry is broken locally by the Higgs condensate itself, and the implication that there is no classical radiation 
at all in the limit $\eta \to 0$ is puzzling.  We leave this issue for future consideration, and focus on the emission of Higgs through the non-perturbative process arising from the self-interaction of the Higgs condensate (see Fig.1).

There is also a quadratic interaction \cite{Srednicki:1986xg}, which gives rise to two-particle emission. The emission rate was recently recalculated and found to be much larger \cite{Long:2014mxa}, and we will see that a small region of parameter space is excluded by two-photon emission alone.
%
\subsection{Emission of Higgs from cusps (NG scenario)}
\label{III}
%

Loops of cosmic string generically form cusps, or points where the tangent vector vanishes and an ideal Nambu-Goto string would move at the speed of light \cite{VS,hindmarsh}. The Nambu-Goto description breaks down where the cores of the string overlap, which occurs in a region of size $\sigma_c \sim \sqrt{L/M}$, where $M^{-1}$ is the string width in the rest frame, and $\sqrt{LM}$ is the boost factor at a distance $M^{-1}$ from an ideal cusp \cite{Olum:1998ag, Vachaspati:2009kq,Dufaux:2012np} (see Fig.1).
By similar arguments, one can also identify a region where the Higgs condensate (with width $\mH^{-1}$) overlaps, which extends over the larger range $\Delta\sigma \sim \sqrt{L/\mH}$. 

Where the string core overlaps, a non-perturbative energy-loss process can occur \cite{Brandenberger:1986vj, BlancoPillado:1998bv}, as segments of string with oppositely oriented flux annihilate, leading to the conversion of an amount of energy $\mu\si_c \sim \mu\sqrt{L/M}$ into radiation of the symmetry-breaking scalar and gauge field $A_\mu$. Some of this energy will be converted to Higgs radiation with very high momentum, of order $M\sqrt{LM}$. 

Similarly, the non-linear interactions of the Higgs condensate
in its overlap region could lead to conversion of a significant fraction of the energy in the Higgs condensate into radiation. One can estimate the available energy, in the case where the condensate is of order $M$, to be 
\ben
E_c \sim M^2 \Delta \si \sim M^2 \sqrt{L/\mH},
\een
and that the subsequent radiation is concentrated around the wave number $m_h\sqrt{\mH L}$.

%
%
\begin{figure}[!htb]
\label{figure1}
\begin{center}
\includegraphics[width=0.4\textwidth]{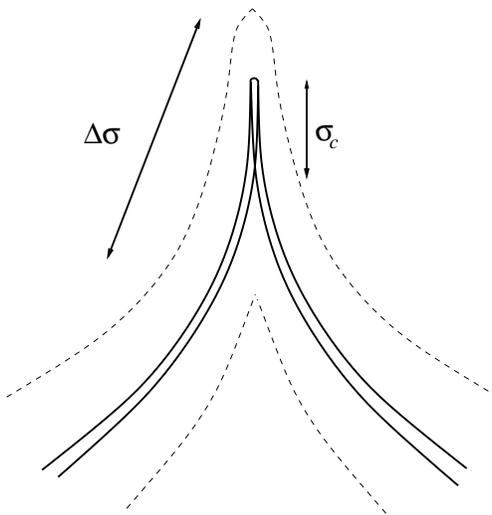}
\caption{Diagram of cusp annihilation, showing the core of the string (solid lines) overlapping over a length $\sigma_c \sim \sqrt{L/M}$, with the Higgs condensate (dashed) interacting over a length $\Delta\sigma \sim \sqrt{L/\mH}$. 
 }
\end{center}
\end{figure}
%
Therefore, given that the loop oscillates with a frequency $L$,  the total power in Higgs emission is 
\begin{equation}
P_h=\frac{\beta_{c}^2\mu}{\sqrt{Lm_h}}, 
\label{GF}
\end{equation}
where $\beta_c^2$ is a numerical factor parametrising the efficiency of the non-perturbative cusp emission process.  

Two-particle emission was recently shown to give a power $P_{hh} \sim M^2/\sqrt{ML}$ \cite{Long:2014mxa}, which is suppressed relative to the non-perturbative process by a factor $\sqrt{\vEW/M}$. We do not study it in detail here, beyond a brief check on the constraints from two-particle emission in the next Section.
%
\subsection{Comparison of Higgs emission with gravitational wave production (NG scenario)}
%
In the NG scenario, the other important decay channel for string loops is the emission of gravitational waves, which are radiated with the power \cite{VS}
\begin{equation}
P_g=\Gamma G\mu^2,
\label{eq:29}
\end{equation}
where $\Gamma\sim 50$. The length $L_e$ at which loops emit the same amount of energy in Higgs particles and gravitational radiation is obtained when $P_h=P_g$, or when the loop has length
\begin{equation}
L_e=\frac{\beta_c^4}{(\Gamma G\mu)^2 m_h}.
\label{eq:30}
\end{equation}
For $L<L_e$, particle emission dominates over gravitational radiation.

Loops of energy $E=\mu L$, radiating gravitational waves and Higgs particles, shrink according to the rate
\begin{equation}
\mu\frac{dL}{dt}=-\Gamma G\mu^2-\frac{\beta_c^2\mu}{\sqrt{Lm_h}}.
\label{eq:31}
\end{equation}
For small loops, Higgs radiation is the dominant energy loss mechanism, and the lifetime of a loop is obtained from (\ref{eq:31}) as
\begin{equation}
\Delta t\sim \frac{L(Lm_h)^{\frac{1}{2}}}{\beta_c^2}.
\label{eq:33}
\end{equation}
Only loops with a lifetime $\Delta t$ greater than the age of the universe $t$ will persist. 
Hence, for a loop of length $L$ to survive to time $t$ it is necessary that $L\gtrsim L_h(t)$, where
\begin{equation}
L_h(t)=\beta_c^{\frac{4}{3}}(m_ht)^{\frac{2}{3}}m_h^{-1},
\label{eq:34}
\end{equation}
for the case of Higgs emission. 

Similarly, the lifetime of a loop when the main energy loss mechanism is by emission of gravitational wave is obtained from (\ref{eq:31}) as
\begin{equation}
\Delta t\sim \frac{L}{\Gamma G\mu}.
\label{eq:35}
\end{equation}
In this case, for a loop of length $L$ to survive to time $t$ it is necessary that $L\gtrsim L_g(t)$, where
\begin{equation}
L_g(t)=\Gamma G\mu t.
\label{eq:36}
\end{equation}
The time at which $L_g(t)$ and $L_h(t)$ are equal is 
\begin{equation}
t_e=\frac{\beta_c^4}{(\Gamma G\mu)^3m_h}.
\label{eq:38}
\end{equation}
The time $t_e$ is also associated with the length $L_e$. Thus, gravitational radiation 
dominates for times $t\gtrsim t_e$ while Higgs radiation dominates for times $t\lesssim t_e$. 

Combining the two sources of energy loss, we find that at time $t$, loops will be longer than a minimum  
length $L_H(t)$ \cite{Dufaux:2012np} given by 
\begin{equation}
L_H(t)=\left\{ \begin{array}{l} L_g(t)=\Gamma G\mu t \,,\qquad\qquad\,\,\,\,\mathrm{for}\,\,\,t\gtrsim t_e\,,\\
L_h(t)=\beta_c^{\frac{4}{3}}(m_ht)^{\frac{2}{3}}m_h^{-1}\,,\,\,\,\,\,\mathrm{for}\,\,\,t\lesssim t_e\,.
\end{array}\right.
\label{eq:37}
\end{equation}
%
\subsection{Higgs emission in the FT scenario}
%
Direct numerical simulation of the Abelian Higgs model, the canonical field theory with a cosmic string solution, shows that there is a non-perturbative radiation mechanism which efficiently turns the energy in string into massive radiation of the fields from which it is made.  

The mechanism allows the string network to evolve in a self-similar manner, known as scaling.  When strings reach scaling, the average distance between strings $\xi$, defined in Eq.\ (\ref{e:IntStrDis}), increases in proportion to the horizon distance.
Loops are produced with an average size which also increases in proportion to the horizon distance.
Scaling behaviour is observed in FT strings for a variety of initial conditions \cite{Vincent:1997cx,Moore:2001px,Bevis:2006mj,Bevis:2010gj}, for inter-string separations of up to the maximum accessible, about 85 in Ref. \cite{Bevis:2010gj}  \footnote{Recent simulations have shown that scaling lasts until at least 250 in the ratio of the string separation to the string width (Daverio, Hindmarsh, Kunz, Lizarraga, Urrestilla, unpublished).}.

In the FT scenario, the efficiency of massive radiation means that a loop of length $L$ 
survives for a time of order $L$ \cite{Hindmarsh:2008dw}, and so the power in massive radiation of a loop is of order $\mu$, much greater than either cusp emission or gravitational radiation.  There is also significant direct emission from the long strings.  It is an interesting and not yet fully-solved problem how the field energy of a loop of size $L$ is transformed into radiation with frequency $M$: in broad outline it involves the coupling of small-scale waves on string to the large-scale modes at cusps \cite{Hindmarsh:2008dw}. The fundamental assumption of the FT scenario is that the massive radiation mechanism continues to operate at all times, and that we can extrapolate the results of the numerical simulations until today.

The net result is that all the energy in the string network is converted to massive radiation, and where there is a coupling between the Higgs and the string fields, we can expect a proportion to appear as Higgs radiation. 
In keeping with our parametrisation of the NG scenario, we will write this proportion as $\bFT^2$. Rather than consider individual loops, we will consider the total power per unit volume of the network itself, $Q_h$. By covariant energy conservation this can be written
\ben
Q_h = - \be_\text{ft}^2\left[ \dot \rho_s + 3H(1+w_s)\rho_s \right],
\een
where $H$ is the Hubble parameter, $\rho_s$ is the total energy density of strings, and $w_s$ is their average equation of state parameter.
%
\section{Evolution of strings and the loop distribution}
\label{s:EvoStrLooDis}
\subsection{Long strings}
Strings are formed in a tangled network, with most of the string length in the form of one infinite string, and 
the rest in a scale invariant distribution of loops \cite{VS,hindmarsh}.  Initially, the strings interact strongly with the 
cosmic fluid and are heavily damped. 
This friction-dominated era lasts until 
the fluid density is sufficiently low that 
strings move freely, apart from Hubble damping (see Section \ref{ss:FriDomEpo}). 
After the end of the friction-dominated era,  
the inter-string distance is proportional to the horizon distance, or $\xi \propto d_H(t).$ 
The interstring distance is defined such that the infinite string density $\rhoInf$ is given by
\ben
\label{e:IntStrDis}
\rhoInf = \frac{\mu}{\xi^2}.
\een
Infinite strings lose energy by a mixture of direct particle production and the formation of loops, which subsequently oscillate and 
decay. In the FT scenario, both mechanisms are important: in the NG scenario only loop production is important.  The loops 
subsequently oscillate and decay either by particle production (FT) or gravitational radiation (NG).
%
\subsection{Loop Distribution}
%
In the NG scenario, loops decay slowly, and comprise most of the string energy density. Therefore in order to calculate the emission of radiation it is necessary to know the loop size distribution, which in turn requires the typical length loops are born with, as well as accurate calculations of how they decay. 
Recent numerical simulations \cite{Ringeval:2005kr,BlancoPillado:2011dq} appear to be converging on a picture in which stable (i.e.\ non-self-intersecting) loops are born with a wide distribution of sizes up to a maximum of 
\begin{equation}
L_i\simeq\beta t_i,
\label{eq:41}
\end{equation}
with $\beta\simeq 0.1$.

Regardless the precise distribution of loops at formation, the number density of loops  with lengths between $L$ and $L+dL$ is in the radiation era \cite{VS}
\begin{equation}
n(L,t)dL \simeq  \nu t^{-\frac{3}{2}} L^{-\frac{5}{2}}dL,\,\,\,\,\mathrm{for} \;\;t<t_{\mathrm{eq}},
\label{eq:42}
\end{equation}
where $L_H(t)\lesssim L\lesssim \beta t_i$, and $\nu \simeq 0.2$ \cite{Blanco-Pillado:2013qja}. Note that $L$ here means the invariant rest length of the string, defined such that the rest energy is $\mu L$. Note also that the number density of loops in Eq. (\ref{eq:42}) is dominated by loops with the lower length $L_H(t)$.

As we are also interested in constraints from observations of the DGRB we need to know the distribution of loops in the matter-dominated era. 
We first note that right at the beginning of the matter era, all loops will have been formed in the radiation era, and will just have the same $L$ dependence as (\ref{eq:42}). The largest loops at that time will have size $L\simeq\beta t_\text{eq}$. As the matter era progresses, new small loops will be produced.  
In recent numerical simulations of NG strings \cite{Blanco-Pillado:2013qja}, the density of loops with $L \ll \beta t_\text{eq}$ produced in the matter era is subdominant when compared with those produced in the radiation era.  We shall assume this property for our loop distribution, while noting that if significant loop fragmentation \cite{Scherrer:1989ha} were to occur, a different loop distribution may be required.

Thus, for loops with sizes in the range $L_H(t)\lesssim L\lesssim \beta t_{\mathrm{eq}}$, 
the number density distribution is 
%
\begin{equation}
n(L,t)dL\simeq \nu t_{\mathrm{eq}}^{\frac{1}{2}} t^{-2}L^{-\frac{5}{2}}dL,\,\,\,\,\mathrm{for} \;\;t>t_{\mathrm{eq}}.
\label{eq:43}
\end{equation}
Indeed, following the results of Ref. \cite{Blanco-Pillado:2013qja}, even today most loops were born in the radiation era for the values of $G\mu$ and $\be_c$ relevant here. As most loops have sizes around the lower cut-off $L_H$, 
a necessary condition for there to exist today loops created in the radiation era is $L_H(t_0)\lesssim \beta t_{\mathrm{eq}}$, which can be translated to $G\mu\lesssim 10^{-8}$ if gravitational radiation dominates or  $\be_c\lesssim 2\times 10^6$ if Higgs radiation dominates. Therefore we take Eq. (\ref{eq:43}) to account only for these predominant loops in the matter era, surviving from the radiation era.

%
\subsection{Friction-dominated epoch}
%
\label{ss:FriDomEpo}
As mentioned earlier, the formation of cosmic strings occurs at very early times, when the universe is dominated by a high density of radiation. In this epoch, the main energy loss mechanism is by friction due to the interaction between the strings and the hot plasma that fills the universe. As the temperature of the universe decreases the strings start to reach relativistic velocities and the friction becomes subdominant. For the model considered in the Appendix, the scattering cross-section for the Higgs particle per unit length of string is roughly\footnote{We are grateful to Andrew Long for pointing this out to us.} $\sigma\sim m_h^{-1}$. 

The average drag force per unit length is approximately $\rho\si$, which drops below the average force due to the string tension $\mu/\xi$
at
\begin{equation}
t_{d}\sim \frac{m_h^{-1}}{G\mu}.
\label{eq:44}
\end{equation}
It is only after this time that the distribution of loops takes the scaling form (\ref{eq:42}). As we are considering constraints from BBN between $10^{-1}$\,s and $10^{12}$\,s (see Sec. \ref{IV}), it is sufficient to have the condition $t_d<10^{-1}$\,s. This implies 
\begin{equation}
G\mu>\frac{m_P}{m_h}\left(\frac{t_P}{10^{-1}\rm{s}}\right)\simeq 5\times 10^{-26}.
\label{eq:44.1}
\end{equation}
Therefore, we do not need to worry about the friction-dominated epoch as long as the BBN and DGRB constraints are inside of the range of applicability given by (\ref{eq:44.1}).  It will turn out that this is indeed the case.
%
\section{COSMOLOGICAL CONSTRAINTS}
\label{IV}
\subsection{Energy density injection (NG scenario)}
%
The total energy density injection rate in Higgs particles emitted by loops is given by \cite{Dufaux:2012np}
\begin{equation}
Q_h(t)=\int dLn_L(L,t)P_h,
\label{eq:46}
\end{equation}
with $P_h$ being the power (\ref{GF}) emitted by a loop of length $L$ and $n_L(L,t)dL$ is given by either (\ref{eq:42}) or (\ref{eq:43}). In the radiation-dominated era, i.e
$t<t_{\mathrm{eq}}$, the integral (\ref{eq:46}) can be found as
\begin{equation}
Q_h^{(\mathrm{r})}(t)\simeq \frac{\bar{\gamma}_h}{(\Gamma G\mu)^{\frac{1}{2}}}\frac{\mu}{t^3}H^{(\mathrm{r})}\left(\frac{t}{t_e}\right),
\label{eq:47}
\end{equation}
where $\bar{\gamma}_h\simeq\frac{\nu}{2}$ and
\begin{equation}
H^{(\mathrm{r})}\left(\frac{t}{t_e}\right)=\left\{ \begin{array}{l}\left(\frac{t}{t_e}\right)^{-\frac{1}{2}},\,\,\mathrm{for}\,\, t>t_e\,,\\
\left(\frac{t}{t_e}\right)^{\frac{1}{6}},\,\,\,\,\,\,\mathrm{for}\,\, t<t_e\,.
\end{array}\right.
\label{eq:47.1}
\end{equation}
In the matter-dominated era, i.e $t>t_{\mathrm{eq}}$, the integral (\ref{eq:46}) gives 
\begin{equation}
Q_h^{(\mathrm{m})}(t)\simeq\frac{\bar{\gamma}_h}{(\Gamma G\mu)^{\frac{1}{2}}}\left(\frac{t_{\mathrm{eq}}}{t_e}\right)^{\frac{1}{2}}\frac{\mu}{t^3}H^{(\mathrm{m})}\left(\frac{t}{t_e}\right),
\label{eq:48}
\end{equation}
with
\begin{equation}
H^{(\mathrm{m})}\left(\frac{t}{t_e}\right)=\left\{ \begin{array}{l}\left(\frac{t}{t_e}\right)^{-1},\,\,\mathrm{for}\,\, t>t_e\,,\\
\left(\frac{t}{t_e}\right)^{-\frac{1}{3}},\,\mathrm{for}\,\, t<t_e\,.
\end{array}\right.
\label{eq:48.1}
\end{equation}
Note that both expressions (\ref{eq:47}) and (\ref{eq:48}) are dominated by loops of the minimum size $L_H(t)$. For early times ($t < t_e$) loops are mainly decaying by Higgs emission, and so $L_H(t) = L_h(t)$, while at late times gravitational radiation is the main decay channel, and so $L_H(t) = L_g(t)$.

We will apply the BBN bounds derived in Ref.  \cite{Kawasaki:2004qu} and summarised in Table \ref{table1}. 
The bounds in \cite{Kawasaki:2004qu} are expressed in terms of $E_{\mathrm{vis}}Y_X(t)$, where $Y_X(t)$ is the yield at time $t$ of a new species $X$ injected into the cosmic medium with average energy  $E_{\mathrm{vis}}$, and subsequently decaying into ``visible'' (i.e. not weakly interacting) states. 
We therefore need the energy density injected in one cosmic time $t$,  in units of the entropy density $s$.
Defining $\Delta\rho_h(t)=tQ_h^{(\mathrm{r})}(t)$, we obtain from (\ref{eq:47}) that  
\begin{equation}
E_{\mathrm{vis}}Y_X(t)=\frac{\Delta\rho_h(t)}{s(t)}\simeq 7.8\bar{\gamma}_h\frac{(t_P\mu)}{(\Gamma G\mu)^{\frac{1}{2}}}\left(\frac{t_P}{t}\right)^{\frac{1}{2}}H^{(\mathrm{r})}\left(\frac{t}{t_e}\right),
\label{eq:49}
\end{equation}
where $s(t)=0.0725 \mathcal{N}^{\frac{1}{4}}(m_P/t)^{\frac{3}{2}}$ is the entropy density, with $\mathcal{N}\sim 10$ being the effective number of degrees of freedom during BBN.

\begin{table}[ht]
\begin{center}
\vspace{3mm} 
\begin{tabular}{|c|c|c|c|}
\hline
$t_s=t/{\rm s}$ & Element & Low bound & High bound  \\\hline 
$1$ & $^4$He & $E_{\mathrm{vis}}Y_X \lesssim 4\times10^{-12}$ GeV\;\;\;\;\;\; &  $E_{\mathrm{vis}}Y_X \lesssim 8\times10^{-11}$ GeV\;\;\\\hline
$3.2$ & $^4$He & $E_{\mathrm{vis}}Y_X\lesssim 2\times {10^{-12}}$ GeV\;\;\;\;\; &  $E_{\mathrm{vis}}Y_X \lesssim 5\times10^{-11}$ GeV\\\hline 
$10$ & $^4$He & $E_{\mathrm{vis}}Y_X\lesssim {1\times10^{-12}}$ GeV\;\;\;\;\; &  $E_{\mathrm{vis}}Y_X \lesssim 3\times10^{-11}$ GeV\;\;\\\hline
$4\times 10^2$ & D & $E_{\mathrm{vis}}Y_X \lesssim 1\times10^{-13}$ GeV\;\;\;\;\; &  $E_{\mathrm{vis}}Y_X \lesssim 3\times10^{-13}$ GeV\;\;\\\hline
$1 \times 10^3$ & D & $E_{\mathrm{vis}}Y_X\lesssim 4\times {10^{-14}}$ GeV\;\;\; &  $E_{\mathrm{vis}}Y_X \lesssim 1\times 10^{-13}$ GeV\;\;\\\hline 
$3\times 10^3$ & D & $E_{\mathrm{vis}}Y_X\lesssim 2\times {10^{-14}}$ GeV\;\;\;\;\;\; &  $E_{\mathrm{vis}}Y_X \lesssim 6\times10^{-14}$ GeV\;\;\\\hline
\end{tabular}
\caption{This table shows the observational BBN \cite{Kawasaki:2004qu} bounds and the correspondent time $t_s$ (in seconds) when they are applicable. From all bounds analysed in  \cite{Kawasaki:2004qu}, the ones that provide the strongest constraints are the low and high Helium  and Deuterium bounds.}
\label{table1}
\end{center}
\end{table}

Will also apply the bound on energy injection in the form of $\ga$-rays derived in \cite{Berezinsky:2010xa} from recent measurements of the DGRB at GeV-scale energies by Fermi-LAT \cite{Abdo:2010nz}. This bound is given by
\begin{equation}
\omega_{\rm{em}}\lesssim 5.8\times 10^{-7}\;\rm{eV/cm}^3.
\label{DGRB2}
\end{equation}
The bound above is quoted in terms of $\omega_{\rm{em}}$, the total electromagnetic energy injected since the universe became transparent to GeV $\ga$-rays, which was at about $t_c \simeq 10^{15}\;\text{s}$.  

Once Higgs particles are emitted they will decay and a significant fraction of the energy cascades into $\ga$-rays, and so Higgs emission is subject to the DGRB bound. 
The electromagnetic energy density from  
Higgs particles decaying into photons can be calculated as
\begin{equation}
\omega_{\rm{em}}=f_{\mathrm{em}}\int_{t_c}^{t_0} dt\frac{Q^{(\rm m)}_h(t)}{(1+z)^4},
\label{eq:52}
\end{equation}
where we choose $f_{\mathrm{em}}=1$ and $(1+z)=(t_0/t)^{2/3}$ in the matter-dominated era. The factor $1/(1+z)^4$ comes from the redshift 
of the photon energy density from the time of production until today. 
The integral (\ref{eq:52}) gives
\begin{equation}
\omega_{\rm{em}}\simeq \frac{3\bar{\gamma}_h}{(\Gamma G\mu)^{\frac{1}{2}}}\left(\frac{t_{\mathrm{eq}}}{t_e}\right)^{\frac{1}{2}}\left(\frac{t_e}{t_0}\right)^{\frac{2}{3}}\frac{\mu}{t_0^2}M(t_c,t_e,t_0),
\label{eq:53}
\end{equation}
where
\begin{equation}
M(t_c,t_e,t_0)=\left\{ \begin{array}{l}\,\,\,\,\,\,\,\,\,\, \left(\frac{t_e}{t_c}\right)^{\frac{1}{3}}-\left(\frac{t_e}{t_0}\right)^{\frac{1}{3}},\,\,\,\,\mathrm{for}\,\,\, t_e<t_c\,, \\
2-\left(\frac{t_c}{t_e}\right)^{\frac{1}{3}}-\left(\frac{t_e}{t_0}\right)^{\frac{1}{3}},\,\,\,\,\mathrm{for}\,\,\,t_c<t_e<t_0\,,\\
 \,\,\,\,\,\,\,\,\,\,\left(\frac{t_0}{t_e}\right)^{\frac{1}{3}}-\left(\frac{t_c}{t_e}\right)^{\frac{1}{3}},\,\,\,\,\mathrm{for}\,\,\, t_e>t_0\,.
\end{array}\right.
\label{eq:54}
\end{equation}
Note that there are three cases to consider, depending on when $t_e$ occurs in relation to $t_c$ and $t_0$. The first case is when $t_e < t_{c}$, 
in which case the contribution to gamma-rays comes from loops decaying mainly into gravitational radiation, most of which are 
of size $L_{g}(t)$.  This contribution, in the top of Eq. (\ref{eq:54}), is dominated by the first term. The second case is when $t_{c}<t_e<t_0$ and the integration (\ref{eq:52}) has two contributions: $t_c\lesssim t\lesssim t_e$, when loops are mostly emitting Higgs particles and are of size $L_h(t)$, and $t_e\lesssim t\lesssim t_0$ when the loops are mostly decaying into gravitational radiation, but contribute also to gamma-rays, and mostly have size $L_g(t)$. The sum of these two contributions (the middle expression in Eq. (\ref{eq:54})) is dominated by the first term in the expression on the right hand side. The third and last case is when $t_0 < t_e$, in which case loops are still decaying predominately into Higgs particles, and are mostly of size $L_h(t)$. In this case, the expression is dominated by the first term in the bottom of Eq. (\ref{eq:54}).

%
\subsection{Constraints on $G\mu$ and $\be_c$ (NG scenario)}
%
Now that we have computed the energy injection by decaying Higgs from cosmic strings in the NG scenario, we can derive the bounds on the string tension $G\mu$ and the efficiency parameter $\be_c$. 
%
\begin{table}[ht]
\begin{center}
\vspace{3mm} 
\begin{tabular}{|c|c|c|}
\hline
$\mathrm{Constraints}$ $\mathrm{on}$ $G\mu$ & Observational bounds & Time interval\\
\hline 
$G\mu\gtrsim 8\times 10^{-10}\beta_c^2$ & \, \raisebox{-0.4cm}{High $^4$He bound} &  \raisebox{-0.4cm}{$1\ {\rm s}\lesssim t_{{\rm BBN}}\lesssim 10\ {\rm s}$}\\
$G\mu\lesssim 1\times 10^{-12}\beta_c^{\frac{2}{3}}$ &  & \\\hline
$G\mu\gtrsim 2\times 10^{-8}\beta_c^2$\;\; & \raisebox{-0.4cm}{Low $^4$He bound} & \raisebox{-0.4cm}{$1\ {\rm s}\lesssim t_{{\rm BBN}}\lesssim 10\ {\rm s}$}\\
$G\mu\lesssim 4\times 10^{-14}\beta_c^{\frac{2}{3}}$ &  & \\\hline
$G\mu\gtrsim 6\times 10^{-10}\beta_c^2$\; & \, \raisebox{-0.4cm}{High D/H bound} &  \raisebox{-0.4cm}{$4\times10^2\ {\rm s}\lesssim t_{{\rm BBN}}\lesssim 3\times 10^3\ {\rm s}$}\\
$G\mu\lesssim 2\times 10^{-14}\beta_c^{\frac{2}{3}}$ &  & \\\hline
$G\mu\gtrsim 2\times 10^{-9}\beta_c^2$\;\; & \raisebox{-0.4cm}{Low D/H bound} & \raisebox{-0.4cm}{$4\times10^2\ {\rm s}\lesssim t_{{\rm BBN}}\lesssim 3\times 10^3\ {\rm s}$}\\
$G\mu\lesssim 6\times 10^{-15}\beta_c^{\frac{2}{3}}$ &  & \\\hline
\end{tabular}
\caption{Constraints on the string tension $G \mu$ in the NG scenario, derived from comparing the predicted energy injection per unit entropy (\ref{eq:49}) with the BBN constraints of Table \ref{table1}, and
expressed as a function of the Higgs emission efficiency parameter $\be_c$.  
As explained in the text, a constraint on $E_{\mathrm{vis}}Y_X(t)$ excludes a wedge-shaped region in the $(G\mu,\be_c)$ parameter space.  The left column shows the equations of the greatest lower bound and the least upper bound resulting from all such regions in the time interval given in the right hand column. 
The union of the excluded regions are plotted in Fig.\ \ref{fig2}. 
%
} 
\label{table2}
\end{center}
\end{table}
%

We first consider the bounds from BBN on energy injection presented in Fig.\ 38 of Ref.\ \cite{Kawasaki:2004qu}.
The bounds are calculated for a generic bosonic $X$ particles with a lifetime around the time of necleosynthesis, 
and a mass $m_X$=100 GeV. This is close to the Higgs mass and, thus, suitable for our estimate.

We have extracted estimated limits on $E_{\mathrm{vis}}Y_X(t)$ at several times (see Table \ref{table1}). From all elements considered in the Fig.\ 38 of Ref. \cite{Kawasaki:2004qu}, the strongest constraints in our case come from the primordial abundances of $^4$He and D. It should be noted that Ref. \cite{Kawasaki:2004qu} takes into account two bounds for both the primordial abundance of $^4$He and D. 
The $^4$He bounds come from abundance estimates from two different groups \cite{Fields:1998gv,Izotov:2003xn}, while the two D bounds come from an estimate of the uncertainty in the measurements of the primordial D abundance in a small number of damped Lyman-$\al$ systems \cite{Kirkman:2003uv}.
We will quote limits derived from the two bounds for both elements, labelling them as low and high bounds. 

Note that the low bound on $E_{\mathrm{vis}}Y_X(t)$ coming from the primordial $^4$He is a factor 20-30 lower than the high bound, whereas the central values of the primordial $^4$He estimates from which the bounds are derived differ by only a few per cent. 
This sensitivity arises because the standard BBN prediction for the $^4$He abundance at the range of baryon-to-photon ratios used by Ref. \cite{Kawasaki:2004qu} is very close to the upper 2$\si$ limit of the low estimate. Hence there is very little room for the extra $^4$He which the energy injection brings about.  

The resulting bounds on $G\mu$ and $\beta_c$ are calculated from Eq.\ (\ref{eq:49}).  The energy injection at the time $t_\text{BBN}$, when viewed as a function of $G\mu$ and $\beta_c$, takes the form of a ridge along the line $t_e(G\mu,\be_c) = t_\text{BBN}$.  Hence, For $t_e>t_\text{BBN}$, there is an upper bound on the combination $G\mu\be_c^{-2/3}$, while for $t_e < t_\text{BBN}$ there is a lower bound on $G\mu\be_c^{-2}$. For a given $t_\text{BBN}$, the excluded region takes the form of a wedge around the line $t_e(G\mu,\be_c) = t_\text{BBN}$.
The numerical values of the strongest bounds on these parameter combinations are given in Table \ref{table2}, along with the source of the bound, and the range of $t_\text{BBN}$ over which the bound is important.

%
\begin{figure}[!htb]
\begin{center}
\begin{tabular}{cc}
\epsfysize=7cm \epsfbox{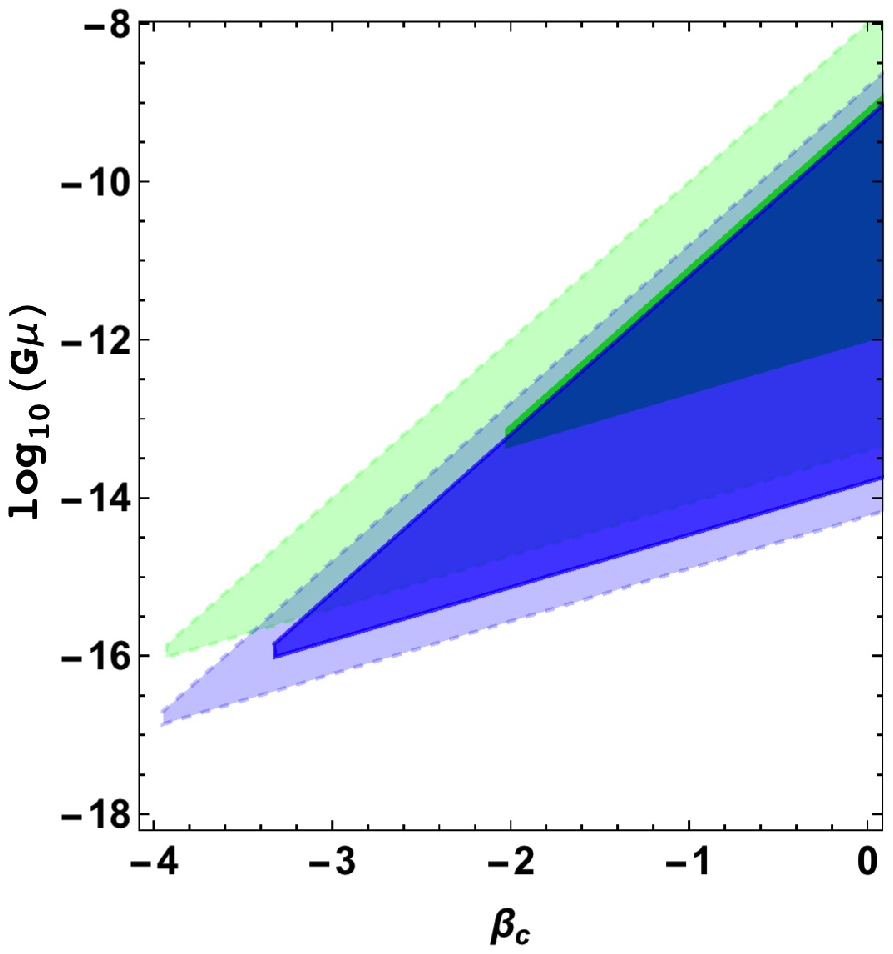}&
\epsfysize=7cm \epsfbox{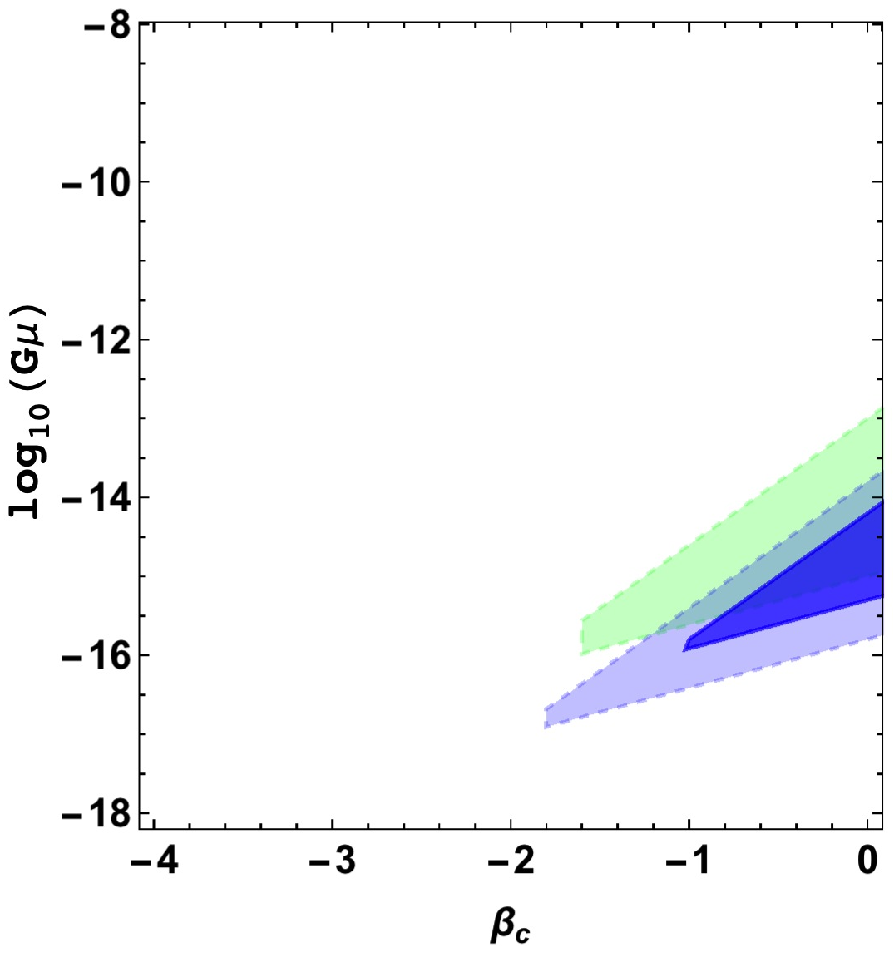}
\end{tabular}
\caption{Constraints on cosmic string tension $G\mu$, in the NG scenario,  derived from 
plotted in the $(\beta_c, G\mu)$-plane for $\nu= 0.2$ and $\Gamma= 50$. The plot on the left shows constraints due to non-perturbative emission of Higgs at cusps, and on the right due to two-particle emission. 
In both plots, green represents constraints due to $^4$He and blue those due to D, while light and dark colours represent bounds from low and high estimates of the abundances.
%
%
Note that there is no bound due to two-Higgs emission from the high $^4$He abundance estimate for physical values $\beta_c < 1$.
}
\label{fig2}
\end{center}
\end{figure}

We plot the excluded regions
in the $(\beta_c,G\mu)$-plane in Fig.\ \ref{fig2}  (left).  
Note that in the case of the $^4$He constraints, a range of $t_\text{BBN}$ contributes significantly, and we see the union of a set of wedge-shaped regions.
We have also checked that there are relevant constraints from the two-Higgs emission and plotted them also in Fig. \ref{fig2} (right).

In both plots, green represents constraints due to $^4$He and blue those due to D, while light and dark colours represent bounds from low and high estimates of the abundances.
The $^4$He and D bounds we have taken are derived, respectively, from the the intervals $1\ {\rm s}\lesssim t_{{\rm BBN}}\lesssim 10\ {\rm s}$ and  $4\times10^2\ {\rm s}\lesssim t_{{\rm BBN}}\lesssim 3\times 10^3\ {\rm s}$. In all cases the excluded values of $G\mu$ are sufficiently large that the  friction-dominated epoch is long finished by the time of BBN (see Eq.\ \ref{eq:44.1}).

We recall that the two-Higgs emission is suppressed over the non-perturbative cusp emission by a factor $\sqrt{\vEW/M}$. As a consequence, on the right plot in Fig.\ \ref{fig2}, one can see that the excluded regions due to $^4$He and D are much smaller, compared with the plot on the left. 
For the physical values $\be_c\lesssim 1$, there is no constraint due to two-Higgs emission from the high $^4$He abundance estimate.

It should be noted that the values for the primordial abundances used in Ref.\ \cite{Kawasaki:2004qu} have undergone revisions and the suggested current values for $^4$He and D can be found in \cite{PDG}. Compared with the abundance values considered in \cite{Kawasaki:2004qu},
the current Helium abundance is about $2\%$ above our high $^4$He value, while the current Deuterium value is about $10\%$ below our low D value, with a significantly reduced uncertainty. It is beyond the scope of this paper to recalculate the bounds on $E_{\mathrm{vis}}Y_X(t)$, but one can estimate that they will change in proportion to the change in the central value of the abundance.  The reduction in the observational errors would make little difference, as the error budget in \cite{Kawasaki:2004qu} is dominated by nuclear cross-sections and hadronic decays.  Hence we argue that a recalculation using \cite{Kawasaki:2004qu} with current abundance estimates would produce bounds on $G\mu$ and $\be_c$ similar to the high $^4$He low D values in Table \ref{table2}, which are quoted to one significant figure only.

Bounds on the energy injection $\omega_{\rm{em}}$ 
from Fermi-LAT diffuse gamma-ray data \cite{Abdo:2010nz,Berezinsky:2010xa} 
(see Eq. \ref{DGRB2}) 
follow from Eqs. (\ref{eq:53}) and (\ref{eq:54}). 
The analytical expressions for the resulting constraints on $G\mu$ and $\be_c$ are found to be 
\begin{equation}
\begin{array}{l} G\mu\gtrsim 2\times 10^{-17}\beta_c^2\,,\,\,\,\,\mathrm{for}\,\,\,t_e\lesssim t_c\,,\\
G\mu\lesssim 1 \times 10^{-15}\beta_c^{\frac{2}{3}}\,,\,\,\,\,\mathrm{for}\,\,\,t_e\gtrsim t_0\,.
\end{array}
\label{DGRB}
\end{equation}
The expressions above formally provide a constraint  only if $\beta_c$ takes unphysically large values greater than about $10^2$, which is found using the middle expression in Eq. (\ref{eq:54}).
%
\subsection{Constraints on $G\mu$ and $\be_\text{ft}$ (Field Theory scenario)}
%
Field theory simulations \cite{Vincent:1997cx,Bevis:2006mj,Hindmarsh:2008dw} suggest that cosmic strings lose energy during their lifetime mainly by classical radiation of massive fields, corresponding to the emission of particles. Thus,  in terms of the total equation of state parameter $w$, the energy injection rate into observable particles from string decay is given by \cite{Hindmarsh:2013jha,VS}
\begin{equation}
\frac{Q_h}{H\rho}=3\be_\text{ft}^2(w-w_s)\Omega_s,
\label{eq:55}
\end{equation}
where $\Omega_s=\rho_s/\rho$ is the string density parameter, with $\rho$ being the total energy density in the universe. 
We recall that the factor $\be_\text{ft}^2$ parametrises the fraction of the available string energy going into Higgs or other Standard Model particles.

It follows from Eq. (\ref{eq:55}) that we can define the total energy density, in Higgs particles, released in the cosmic medium as 
\begin{equation}
\Delta\rho_h(t)=tQ_h=\bar{\ga}_{h,\text{ft}} \frac{\mu}{t^2},
\label{eq:56}
\end{equation}
%
%
%
\begin{figure}[!htb]
\begin{center}
\begin{tabular}{cc}
\epsfysize=7cm \epsfbox{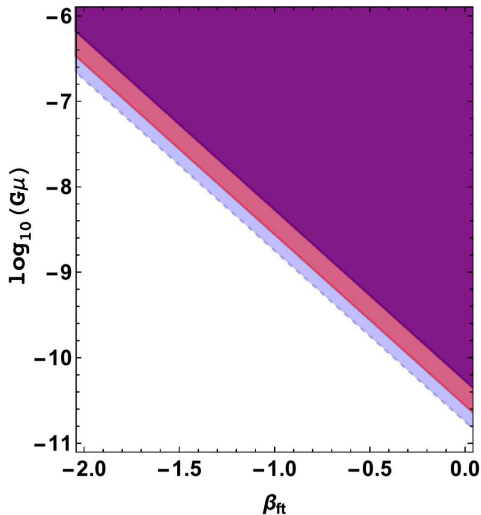}&
\end{tabular}
\caption{Constraints on cosmic string tension $G\mu$, in the FT scenario, derived from BBN \cite{Kawasaki:2004qu} and the DGRB \cite{Berezinsky:2010xa},  plotted in the $(\be_\text{ft}, G\mu)$-plane. Deuterium is the only element that provides constraints in this case. Thereby, the light blue region bounded below by the blue dashed line is excluded by the low D bound. The red region bounded below by the solid red line is excluded by the combination of the low D and DGRB bounds while the purple region bounded below by the solid purple line is excluded by the low D, DGRB and high D bounds.}
\label{fig3}
\end{center}
\end{figure}
%
where $\bar\ga_{h,\text{ft}}=3\be_\text{ft}^2(w-w_s)/x^2$, and we have defined $x=\alpha \sqrt{\mu/\rho_s t^2}$, which is about 0.7 $(\alpha=\sqrt{2})$ in the radiation era and 0.9 $(\alpha=\sqrt{3/2})$ in the matter era \cite{Bevis:2006mj,Hindmarsh:2008dw}. The string equation of state parameter is $w_s \simeq -0.15$ in the matter era and $w_s \simeq -0.13$ in the radiation era.\footnote{D. Daverio, M. Hindmarsh, M. Kunz, J. Lizarraga, and J. Urrestilla, unpublished.}
We therefore take  $\bar\ga_{h,\text{ft}} \simeq 2.8\be_\text{ft}^2$ in the radiation era and $\bar\ga_{h,\text{ft}}\simeq 0.5\be_\text{ft}^2$ in the matter era. 

Using (\ref{eq:56}) we can obtain the total energy density in Higgs particles in the radiation-dominated era in units of entropy density as being 
\ben
\label{e:EneFTBBN}
E_{\rm{vis}}Y_X=\frac{\Delta\rho_h(t)}{s(t)}\simeq 2.2\times 10\be_\text{ft}^2G\mu\left(\frac{m_P}{t}\right)^{\frac{1}{2}}.
\een
We can also obtain the electromagnetic energy density injected by the decay of strings as 
\ben
\label{e:EneFTDGRB}
\omega_{\rm{em}}=f_{\rm{em}}\Delta\rho_h(t)\simeq 0.5\be_\text{ft}^2\frac{\mu}{t^2},
\een 
where $f_{\rm{em}}$ is the fraction of the total Higgs energy (\ref{eq:56}) that ends up in $\ga$-rays in the Fermi-LAT sensitivity range 0.1 - 100 GeV. 
The primary decay channel is $b\bar{b}$, which will produce many photons via pion decays. Photons will also be produced by electromagnetic cascades caused by interactions with the various kinds of background radiation. Hence it is reasonable to take $f_{\rm{em}}$ to be of order unity, and we will take $f_{\rm{em}} = 1$ as an adequate level of modelling.

%
\begin{table}[ht]
\begin{center}
\vspace{3mm} 
\begin{tabular}{|c|c|c|}
\hline
$\mathrm{Constraints}$ $\mathrm{on}$ $G\mu$ & Observational bounds & time\\
\hline 
$G\mu\lesssim 2.7\times 10^{-11}\be_\text{ft}^{-2}$ &  DGRB   &   $t=t_0$\\\hline
$G\mu\lesssim 5.4\times 10^{-11} \be_\text{ft}^{-2}$\;&  \,High D bound &   $t=t_{\mathrm{BBN}}$\\\hline
$G\mu\lesssim 1.8\times 10^{-11} \be_\text{ft}^{-2}$ & \,Low D bound & $t=t_{\mathrm{BBN}}$\\\hline
\end{tabular}
\caption{Constraints on the string tension $G \mu$ in the FT scenario. This table shows the analytical bounds we obtain from observations of DGRB and BBN, for emission 
of Higgs, as a function of the parameter $\be_\text{ft}$.  Note that the strongest constraints come from D at time $t_{\mathrm{BBN}}=3\times10^{3}\,\mathrm{s}$.}
\label{table3}
\end{center}
\end{table}
%

By applying the bounds presented in table \ref{table1} and in Eq. (\ref{DGRB2}) one gets, respectively, the BBN and DGRB constraints presented in table \ref{table3}, for the FT scenario.
The constraints are also plotted in the $(\be_\text{ft}, G\mu)$-plane, as can be seen in Fig.\ref{fig3}. We can notice in the plot that the strongest BBN constraints come from the abundance of Deuterium at $t_{{\rm BBN}}=3\times 10^3\;{\rm s}$. The low D, DGRB and high D bounds are, respectively, bounded below by a light blue dashed line, a solid red line and a solid purple line. Thus, the light blue region is excluded by the low D bound, the red region is excluded by the combination of the low D and DGRB bounds and the purple region is excluded by the low D, DGRB and high D bounds. Note that the low D bound is the strongest one.

We noticed that our DGRB bound is stronger than the one quoted in \cite{Hindmarsh:2011qj}, $G\mu\lesssim 10^{-10} f^{-1}$, where $f = \be_\text{ft}^2$ is the fraction of the strings' energy appearing as standard model particles. The previous limit was based on EGRET data \cite{Sreekumar:1997un} and an older analysis of the cascade into $\ga$-rays by a different group \cite{Bhattacharjee:1998qc}.

%
\section{conclusions}
%
In this paper, we have considered bounds on cosmic string scenarios coming from Big Bang Nucleosynthesis and the Diffuse Gamma-Ray Background in both Nambu-Goto and field theory scenarios of cosmic strings, assuming that the strings have a Higgs condensate of order the cosmic string mass scale $M$.  We  show, by reference to \cite{Haws:1988ax}, that large condensates are to be expected in a wide region of parameter space of models where the Higgs portal coupling is stronger than the self-coupling of the symmetry-breaking scalar field.

In the NG scenario we assumed that the dominant particle production comes from non-perturbative emission from
cusps on string loops, and we also assumed an average of one large cusp per loop per period of oscillation.  The Higgs emission rate in the NG scenario is then (\ref{GF}).  
We derived 
the distribution of loops (\ref{eq:42}) that were born in the radiation-dominated era (relevant for BBN bounds), and the distribution of loops (\ref{eq:43}) that were born in the radiation-dominated era and still survive in the matter-dominated era (relevant for the constraints due to DGRB). 
We checked that the bounds apply to strings which have ceased being friction-dominated using (\ref{eq:44.1}). 

The Higgs energy density injection per expansion time in units of entropy (relevant for the BBN bounds) was presented for the NG scenario in (\ref{eq:49}). The electromagnetic energy density from Higgs particles decaying into photons (relevant for DGRB) was presented in  (\ref{eq:53}).
The BBN and DGRB constraints on the NG scenario parameters (string tension $G\mu$ and Higgs radiation efficiency $\be_{c}$) were presented in table \ref{table2} and Eq. (\ref{DGRB}), respectively. All these constraints are plotted in Fig.\ref{fig2}.  

We also applied the bounds  in Table \ref{table1} to the FT scenario, where 
a fraction of the total energy of the string network of order $\be_\text{ft}^{2}$ goes into Higgs radiation.
The high and low D bounds are, respectively, $G\mu\lesssim 5.4\times10^{-11}\be_\text{ft}^{-2}$ and $G\mu\lesssim 1.8\times 10^{-11}\be_\text{ft}^{-2}$ whereas the DGRB bound is $G\mu\lesssim 2.7\times 10^{-11}\be_\text{ft}^{-2}$. These constraints are plotted in Fig.\ \ref{fig3}.
The DGRB bound is significantly stronger than previous estimates \cite{Hindmarsh:2011qj}, thanks to improved modelling and more recent data from Fermi-LAT \cite{Berezinsky:2010xa}.

An extra spontaneously broken U(1) symmetry, whose symmetry-breaking fields are coupled to the Standard Model via the Higgs portal, is a rather conservative extension of known physics.  It is interesting that there are strong cosmological constraints on such models at high energy scales, complementary to those from accelerator searches.   For the future, 
the uncertainty in the modelling of the Higgs emission, parametrised by the efficiency amplitudes $\beta_\text{c}$ and $\be_\text{ft}$, 
can be reduced by numerical simulations of the Abelian Higgs model with extra fields.

\emph{Note added:} During the refereeing process, another paper on very similar topics appeared \cite{Long:2014lxa}. This paper additionally considers Higgs emission from kinks, and constraints from Cosmic Microwave Background spectral distortions and cosmic rays. The constraints from primordial Helium presented in \cite{Long:2014lxa} for cusp emission in the NG scenario are consistent with ours (the green region on the left plot in Fig.\ \ref{fig2}). Constraints from the primordial Deuterium abundance were not taken into account, and the FT scenario was not considered. 
\acknowledgments
H.F.S.M.\ is grateful to the Institute of Cosmology at Tufts University, where part of this work was done, for the hospitality and to Alexander Vilenkin and his group for helpful discussions. 
We are indebted to Jeffrey Long, Tanmay Vachaspati, Eray Sabancilar and especially Andrew Long for comments on a draft of this article. 
H.F.S.M.\ also would like to thank the Brazilian agency CAPES for financial support. 
MH acknowledges support from the Science and Technology Facilities Council (grant number ST/J000477/1).
%

\appendix 

\section{The Higgs condensate and the string-Higgs coupling}
\label{A2}
%
For analysing the classical string solutions, it is sufficient to consider a subset of the bosonic fields with 
a U(1)$'\times$U(1)$_Z$ theory, corresponding gauge fields $A_\mu$ and $Z_\mu$, coupling constants $(g',g_Z)$, and complex scalar fields $\phi$ and $H$. We interpret $H$ as the lower component of the Higgs doublet $\Phi$.
We work in a diagonal basis for the gauge fields, and assume that the U(1)$'$ symmetry is broken at a scale much higher than the electroweak scale, so that we can neglect coupling of the $\phi$ field to $Z_\mu$.  The scalar field charges can then be written $(1,0)$ and $(q_H,1)$. 

The energy functional for the vortex in 2D (which is the functional for the energy per unit length in 3D) is 
\ben
E = \int d^2y\left( |D_a \phi |^2 + |D_a H |^2 + \half B_A^2 + \half B_Z^2  +V(\phi,H) \right),
\label{B1}
\een
where $d^2y=rdrd\varphi$, $a=(1,2)$ and 
\ben
V(\phi,H) = \half \la_1 (|\phi|^2  - M^2)^2 + \half \la_2 (|H|^2 - \eta^2)^2 +  \la_3(|\phi|^2 - M^2)(|H|^2 - \eta^2),
\label{B2}
\een
is the potential. The covariant derivatives and the two magnetic field strengths are given by
\ben
D_a \phi = (\pa_a - ig'A_a)\phi, \quad D_aH = (\pa_a - ig'q_HA_a - i g_Z Z_a)H,
\label{B3}
\een
and 
\ben
B_A = \ep_{ab}\pa_a A_b, \quad B_Z = \ep_{ab}\pa_a Z_b.
\label{B4}
\een
In terms of the standard SU(2) coupling $g$ and the weak mixing angle $\thW$, $g_Z = g\tan\thW/2$.  
We assume that the ground state is $|\phi| = M$ and $|\Phi| = \eta$, which means that $\la_1\la_2 > \la_3^2$.

An Ansatz for a static cylindrically symmetric solution is
\ben
\phi = M f(x)e^{i\th}, \quad H = M h(x), \quad A_a = \hat\th_a \frac{Ma(x)}{x}, \quad Z_a = \hat\th_a \frac{Mz(x)}{x},
\label{B5}
\een
where $x = g'Mr$. In this case the static energy functional becomes
\ben
\frac{E}{2\pi M^2} = \int dx x \left[ \left(f'\right)^2 + \frac{(1-a)^2}{x^2} f^2   + \left(h'\right)^2 + \frac{( q_H a + z)^2}{x^2} h^2   +      
 \frac{1}{2} \left( \frac{a'}{x}\right)^2 + \frac{1}{2} \al^2 \left( \frac{z'}{x}\right)^2 + \tilV(f,h) \right],
 \label{B6}
\een
where $\al^2={g'}^2/g_Z^2$ and 
\ben
\tilV(f,h) = \half \tilla_1 (f^2  - 1)^2 + \half \tilla_2(h^2-h_\text{ew}^2)^2 +  \tilla_3(f^2 - 1)(h^2 - h_\text{ew}^2),
\label{B8}
\een
with $h_\text{ew}^2=\eta^2/M^2$ and $\tilla_i = \la_i/{g'}^2$. We assume that $h_\text{ew} \ll 1$. Then in this limit, the Higgs mass is $m_h = \sqrt{2\la_2}\eta = \sqrt{\la_2}\vEW$.

The standard Nielsen-Olesen solution is obtained by minimising $E$ subject to the constraints $h=0$ and $z=0$.  We denote this solution $\fNO$, $\aNO$, which has the following properties
\ben
\fNO\simeq
\begin{cases}
	f_0x, & \\
          1-f_1 x^{-1/2}\exp(-\sqrt{\tilla_1} x), & \cr
\end{cases}
\quad
\aNO\simeq
\begin{cases} 
	a_0x^2-{f_0^2} x^{4}/4, & \quad x\to 0;\\
               1-a_1x^{1/2}\exp(-x), & \quad x\to\infty,
\end{cases}
\label{B9}
\een
with $f_0$, $f_1$, $a_0$ and $a_1$ O(1) constants.
In the case $\tilla_1>4$, $x^{-1/2}\exp(-\sqrt{\tilla_1}x)$ is replaced by $x^{-1}\exp(-2x)$.
Roughly speaking, the gauge field $\aNO$ stays close to zero for  $ x \lesssim 1$, while the 
scalar field $\fNO$ stays close to zero for $ x \lesssim X_f = \max(1/\sqrt{\tilla_1},1/2)$. Outside these distances, both $\aNO$ and $\fNO$ approach 1 exponentially.

To show that there is a Higgs condensate, it is sufficient to demonstrate that there is a field configuration with $h= h_0 \ne 0$ at the core of the string which reduces the energy.  The configuration need not be a solution to the field equations: as the solution minimises the energy one is guaranteed that the true solution has even lower energy.

We first note that at distances $ X_f \ll x $, the equation for the Higgs field is
\ben
\label{e:FieEqnH}
-\frac{1}{x} \frac{d}{dx}\left(x \frac{d h}{dx} \right) + \tilla_2 (h^2 - h_\text{ew}^2)h \simeq 0,
\een
which for $h \gg h_\text{ew}$ has the solution 
\ben
\label{e:hSol1}
h \simeq \frac{1}{\sqrt{\tilla_2} x}.
\een
This solution is therefore a good approximation for $x \ll g'M/\mH$, provided it can be matched on to a solution at $ x \ll X_f$.

For  $g'M/\mH \ll x$, the Higgs field relaxes to its vacuum value $h = h_\text{ew}$ according to 
\ben
\label{e:hSol2}
h = h_\text{ew}\left( 1 + \frac{h_1}{\sqrt{r \mH}}\exp(- r \mH )  \right),
\een
where $h_1$ is a constant.

Near the core of the string, at $x \ll 1$, the field equation is  
\ben
\label{e:FieEqnH2}
-\frac{1}{x} \frac{d}{dx}\left(x \frac{d h}{dx} \right) + (\tilla_2 h^2 - \tilla_3)h \simeq 0 ,
\een
which means that $\tilla_2 h_0^2 < \tilla_3$ in order that the condensate be at a maximum at $x=0$.

We accordingly make an ansatz representing a condensate of amplitude $h_0$ and width $X_h$:  
\ben
\label{e:hAns}
\chi_s = 
\begin{cases}
	h_0 &  x \le X_h, \\
	{1}/{\sqrt{\tilla_2} x} & X_h < x, 
\end{cases}
\een
where $h_0 = {1}/{\sqrt{\tilla_2} X_h}$. The amplitude and the width are linked by the field equation (\ref{e:FieEqnH2}), which forces $X_h \gtrsim 1/\sqrt{\tilla_3}$. We set $z=0$ for the ansatz: the field equations will generate some magnetic flux $B_Z$, but this can only reduce the energy further.

The change in the energy due to the Higgs condensate with ansatz (\ref{e:hAns}) is 
\bea
\frac{\De E}{2\pi M^2} &=& 
\int_0^\infty dx x \left[ \left(h'\right)^2 + \frac{( q_H \aNO)^2}{x^2} h^2   +      
 \half \tilla_2 h^4 + \tilla_3 (\fNO^2 - 1) h^2 \right], \nonumber \\
 &\simeq&
h_0^2\left[1+\frac{q_H^2}{2}(1+\ln X_h^2) - \frac{1}{2}C_3(\tilla_i,h_0)\right],
 \label{B15}
\eea
where we have neglected the terms $h_\text{ew}$ in Eq. (\ref{B8}) and the function $C_3(\tilla_i,h_0)$ is 
\ben
\label{e:C3X}
C_3(\tilla_i,h_0) \simeq 
\begin{cases}
          {\tilla_3}X_f^2, & \tilla_3 \ll \tilla_1, \\
	 \tilla_3X_h^2\left[1+ \ln \left( \frac{X_f^2}{X_h^2} \right)\right], & \tilla_3 \gg \tilla_1.
\end{cases}
\een
For the integration in Eq. (\ref{B15}) we assumed that both $X_h$ and $X_f$ are bigger than one. 
The  change (\ref{B15}) must be negative in order for the total energy (\ref{B6}) be reduced. From Eq. (\ref{e:C3X}), and using $X_f=1/\sqrt{\tilla_1}$, we see that $\De E$ cannot be negative for $\tilla_3 \ll \tilla_1$. In the opposite limit $\tilla_3 \gg \tilla_1 $, $\De E$ is made as negative as possible by making $X_h$ as small as possible without violating the condition $X_h \gtrsim 1/\sqrt{\tilla_3}$, so that $h_0^2$ is as large as possible. Hence $X_h \sim 1/\sqrt{\tilla_3}$, and  
\ben
h_0 \sim \sqrt{\frac{\tilla_3}{\tilla_2}}, 
\een  
giving a reduction in energy ${\De E} \sim -{M^2}{\la_3}/{\la_2}$.
Hence, with $ \tilla_3 \gg \tilla_1 $, we can write down an ansatz for the Higgs condensate, whose mass scale is $M$, which reduces the energy by O($M^2$). 

Let us now recall the condition $\la_3^2 < \la_1 \la_2$ which guarantees that $|f| = 1$ and $|h| = \eta/M$ is the global minimum. 
Thus, we see that $\la_3$ is bounded both above and below, and the part of parameter space most favourable for a large condensate while preserving the correct vacuum is 
\ben
\la_1 \ll  \la_3 < \sqrt{\la_1\la_2}. 
\een
Given that $\la_2 \simeq 0.25$ for the measured value $\mH \simeq 126\;\text{GeV}$, the favourable region for a Higgs condensate is 
\ben
\la_1 \ll  \la_3 \lesssim 0.5\sqrt{\la_1}.
\een
We observe that in order for it to be possible to satisfy these inequalities, we must have $\la_1 \ll 1$.

Our ansatz gives only an upper bound on the energy of the true solution, but we already know that it is at least O($M^2$) below that of the Nielsen-Olesen string. We conclude that the true solution must have yet lower energy, which is still O($M^2$), as there is no other relevant scale in the problem.  In order to generate a reduction of this size, the condensate must be of O($M$). We note that a condensate of order $\vEW$ could reduce the energy by only O($\vEW^2$).
\begingroup\raggedright\endgroup

%
\end{document}